\documentclass[twocolumn,pra,superscriptaddress]{revtex4}
\usepackage{ amsmath,amssymb,}
\usepackage{graphicx}% Include figure files
\usepackage{dcolumn}% Align table columns on decimal point
\usepackage{bm}% bold math
\usepackage[colorlinks,citecolor=red]{hyperref}

\usepackage{rotating}
\usepackage{floatrow}

\newcommand{\rmnum}[1]{\romannumeral #1}
\newcommand{\Rmnum}[1]{\expandafter\@slowromancap\romannumeral #1@}

\begin{document}

%\preprint{APS/123-QED}

\title{Nonlinear Bloch-Zener oscillations for Bose-Einstein condensates in a Lieb optical lattice}% Force line breaks with \\

\author{Peng He}
\affiliation{National Laboratory of Solid State Microstructures
and School of Physics, Nanjing University, Nanjing 210093, China}

\author{Zhi Li}
\email{lizhiphys@126.com}
\affiliation{Guangdong Provincial Key Laboratory of Quantum
Engineering and Quantum Materials, SPTE, South China Normal
University, Guangzhou 510006, China}

%\author{Shi-Liang Zhu}
%\email{slzhu@nju.edu.cn}
%\affiliation{National Laboratory of Solid
%State Microstructures and School of Physics, Nanjing University,
%Nanjing 210093, China}

%\affiliation{Guangdong Provincial Key Laboratory of Quantum
%Engineering and Quantum Materials, SPTE, South China Normal
%University, Guangzhou 510006, China}



\date{\today}% It is always \today, today,

\begin{abstract}
We investigate Bloch-Zener oscillations and mean-field Bloch bands of a Bose-Einstein condensate (BEC) in a Lieb optical lattice. We find that the atomic interaction will break the point group symmetry of the system, leading to the destruction of the Dirac cone structure, while the flat band is preserved on the highly symmetric lines. Due to the nonlinear effect, a tubular band structure with a flat band will appear in the system. Furthermore, comparing with that the tight-binding (TB) model fails to describe the interacting bosonic systems in the honeycomb lattice, we show that the TB model is applicable to study the nonlinear energy band structures for the Lieb lattice. In addition, we show that the loop structure can be determined by the observation of the chaos of the state in the Bloch-Zener oscillations.
\end{abstract}

\pacs{Valid PACS appear here}% PACS, the Physics and Astronomy
\maketitle

%\tableofcontents

%\textit{Introduction-}\quad
\section{Introduction}
Two-dimensional condensed matter such as graphene and similar materials~\cite{Liu2011,Mal2012,Zhou2014} with Dirac cones at Fermi energy, in which low-energy excitations at high symmetry points can be described by the Dirac equation, are of great theoretical and experimental interests due to their unique electronic properties. Recent advances in exploring  spin-orbit coupling and artificial gauge fields for neutral atoms~\cite{DWZhang2018AIP,Dali2011,Glodman2014,Galtski2013,SLZhu2006,SLZhu2011} make possible the creation and manipulation of Dirac cones in optical lattices~\cite{Tarr2012,Monta2009,SLZhu2007,DWZhang2012}. Different lattice geometries of one or two dimensions in ultracold atomic gases~\cite{Sol2011,Sol2012,Tarr2012} have been realized experimentally, such as honeycomb lattice with Dirac points at the corner of Brillouin zone (BZ)~\cite{Wun2008}, the Lieb lattice with a flat band \cite{Taie2015}, and the kagome lattice \cite{Jo2012,SZhang2019}, \emph{etc}. Furthermore, some well-known models in condensed matter physics such as Harper-Hofstadter model and Haldane model~\cite{Miv2013,Aidelsburger2013,Aidelsburger2015,Jotzu2014,DWZhang2017,LBShao2008} have been theoretically investigated and experimentally realized  in optical lattices with promising controllability. The band structure of the cold atom systems can be reconstructed by combining techniques of Bloch-Zenor oscillations with adiabatic mapping of cold atoms, and the existence of Dirac points can be revealed by the momentum-resolved inter-band transitions  \cite{Mor2006,Sal2007,Kling2010,Lim2012,Lim2015,DWZhang2016,FMei2012,YQZhu2017}.

On the other hand, with high purity and tunability, the ultracold atomic system is one of the best quantum simulators ~\cite{Lewe2007,DWZhang2018AIP,Bloch2008} to study nonlinear dynamics for a BEC with weak interaction. For instance, the lattice structure can be changed by adjusting the laser intensity, while the interaction can be manipulated by tuning s-wave scattering length~\cite{Bloch2008}. In this paper, a nonlinear Gross-Pitaevskii (GP) theory is derived by applying the mean-field approximation to a weakly coupling system. Besides the atomic systems, GP equation can also be used to study other nonlinear systems such as photonic systems with Kerr nonlinearity, exciton-polariton condensates, acoustic cavities, and circuit resonators~\cite{Hennig1999,Lagoudakis2008,Ley2012,Ley2016,Hadad2016,Whittaker2018,Nejad2019}. Previous studies have shown that the mean-field Bloch bands for a BEC in an optical lattice possess unique loop structures~\cite{Dia2002,Wu2002,Wang2006,Bona2017}. As predicted in Ref.~\cite{Chen2011}, while Dirac points in a honeycomb lattice are broken by the nonlinearity and an intersecting tubular structure appears around the K point, the TB model, however, fails to describe the nonlinear dynamics in this system~\cite{Chen2011}. In this article, we study BEC in an optical Lieb lattice. The Lieb lattice features a diabolic single dirac cone with a flat band, distinguishing from most systems in which the Dirac points come up in pairs (the so-called Fermion doubling) \cite{Shen2010,GSilva2014,JWei2019,Mukherjee2015,Poli2017}. The flat band is attributed to the bipatite nature of the lattice \cite{Mielke1991,ZLiu2014}, leading to exotic physics which is entirely determined by the interaction and topology due to constant dispersion caused by destructive interference \cite{Kopnin2011,JUlku2016}. Thus the nonlinear Lieb lattice may serve as a natural, and possibly indispensable extension of the honeycomb lattice considered in Ref.~\cite{Chen2011}.
Our results show that the nonlinear Dirac cone for the Lied lattice has a similar tubular structure to that of Ref. \cite{Chen2011}. Furthermore, the Dirac points in our model are protected by a plane point group symmetry. Any small interatomic interaction will break the symmetry, then the Dirac points will be lifted. And more importantly, the TB model works well in the vicinity of the M point for our system unexpectedly, which is sharply different with the results in the honeycomb lattice explored in Ref.~\cite{Chen2011}.

%And the flat band exists on high symmetry lines in presence of nonlinearity. Furthermore, the TB model works well in the vicinity of the M point for our system unexpectedly.
The interplay of nonlinear superfluidity and relativistic Dirac dynamics leads to many novel effects, such as adiabatic failure, chaos~\cite{Wu2000,Wang2006} and dynamic instability~\cite{Bro2001}, which can be utilized to reveal information on the nonlinear energy bands. In this article we prove that the Bloch-Zener oscillations for the BEC in an optical lattice can be used as an efficient tool to study the structure of the nonlinear energy bands. The regime of fold bands can be determined by detecting the time evolution of the population of sublattices during Bloch oscillations, while the nonlinearity strength could be measured by the interband transition probability of specific tunneling process.

The paper is organized as follows. In Sec. \ref{sec2}, we numerically calculate the energy spectra from both the GP Hamiltonian and the Bloch Hamiltonian of the TB model. And we discuss how the nonlinearity breaks the Dirac cone from the perspectives of symmetry. In Sec. \ref{sec3}, we consider atoms performing Bloch-Zener oscillations. The transition probabilities for different tunneling process are given both in the adiabatic limit and sudden limit. Finally, a short conclusion is given in Sec. \ref{sec4}.

%%%%%%%%%%%%%%%%%%%%%%%%%
\section{Superfluid in Lieb lattice}\label{sec2}
\begin{figure}[htbp]
	\centering
	\includegraphics[width=0.9\textwidth]{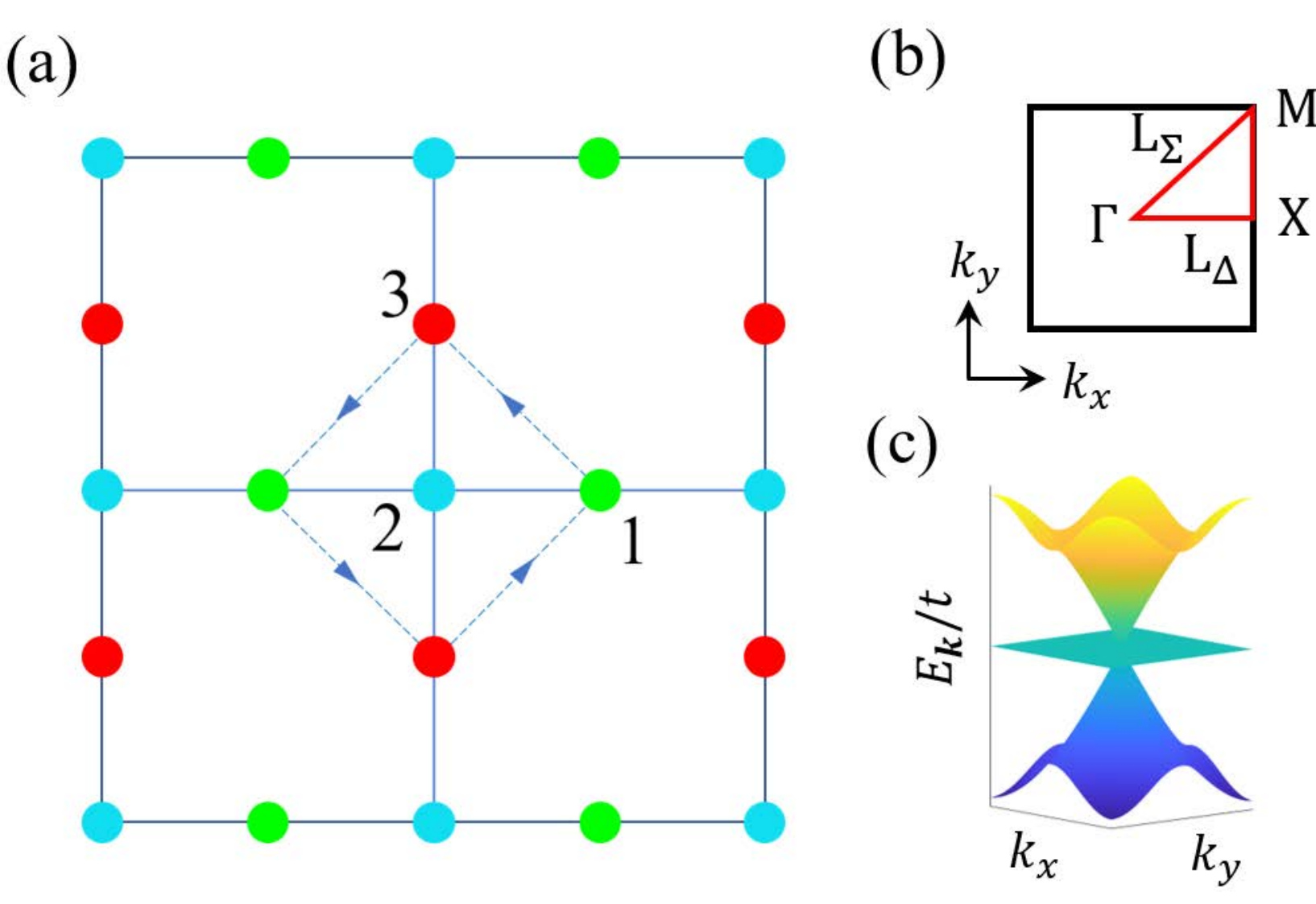}
	\caption{(a) Lattice configuration for Lieb lattice, which has three inequivalent sites in one unit cell. The blue dashed line illustrates next-to-nearest-neighbor hopping, which breaks the sublattice symmetry.  (b)  High symmetric points in the first Brillouin zone. (c) Energy spectra of the tight-binding model in the vicinity of M point for $g=0$.}
	\label{fig1}
\end{figure}

The mean-filed description of a condensate in superfluid phase leads to the Gross-Pitaevskii (GP) equation,
\begin{equation}
i \hbar \frac{\partial \psi}{\partial t}=-\frac{\hbar^{2}}{2 m} \nabla^{2} \psi+V_{latt}(\mathbf{r}) \psi+\frac{4 \pi \hbar^{2} a_{s}}{m}|\psi|^{2} \psi\,,
\label{eq_GP}
\end{equation}
where $m$ is the mass of particle, $a_s$ is the scattering length and $V_{latt}(\mathbf{r})$ is the optical lattice potential of configurations of Lieb lattice (see Fig. \ref{fig1}). The nonlinear strength is denoted as $U\equiv4 \pi \hbar^{2} a_{s}/m$ in following discussions. The lattice  $V_{latt}(\mathbf{r})$ could be experimentally realized by superimposing three pairs of laser beams with the following formation \cite{Taie2015,Shen2010,Weeks2010},

\begin{equation}
\begin{split}&V_{latt}(x, z)=-V_{\mathrm{long}}^{(x)} \cos ^{2}\left(k_{\mathrm{L}} x\right)-V_{\mathrm{long}}^{(y)}\cos ^{2}\left(k_{\mathrm{L}} y\right)\\& -V_{\mathrm{short}}^{(x)} \cos ^{2}\left(2 k_{\mathrm{L}} x+\phi_{x}\right)-V_{\mathrm{short}}^{(y)}\cos ^{2}\left(2 k_{\mathrm{L}} y+\phi_{y}\right) \\& -{V_{\mathrm{diag}}^{(x)} \cos^{2}\left(k_{\mathrm{L}}(x-y)+\varphi\right)}- V_{\mathrm{diag}}^{(y)}\cos ^{2}\left(k_{\mathrm{L}}(x+y)+\varphi\right)\,,\end{split}
\label{eq_pot}
\end{equation}
where $k_L=2\pi/\lambda$ is the wave number of the lattice and $\lambda$ the wavelength of the lasers. The potential amplitudes $V_{\mathrm{long}}^{(x,y)}$, $V_{\mathrm{short}}^{(x,y)}$ and $V_{\mathrm{diag}}^{(x,y)}$ could be tuned by adjusting the laser intensities, $\phi_{x,y}$ and $\varphi$ are the phase of laser beams. For simplicity we choose $(V_{long}^{(x)},V_{short}^{(x)},V_{diag}^{(x)})=
(V_{long}^{(y)},V_{short}^{(y)},V_{diag}^{(y)})=(V_{long},V_{short},V_{diag})$ and $\phi^{(x,z)}=\varphi=0$.

The Landau instability of the BEC superfluid is not much affected by the lattice strength \cite{ZChen2010}. As the lattice strength increases, the system goes into the tight binding regime. Following the methods outlined in Ref. \cite{Smerzi2002,Trombettoni2001,Hu2015}, we map the GP equation (\ref{eq_GP}) onto a  discrete nonlinear Scr$\mathrm{\ddot{o}}$dinger equation to contruct an effective tight-binding Hamiltonian. We write the bosonic field as a sum over the three sub-lattices $\Psi=\sum_{i,\alpha} \Psi_{i,\alpha}u(\mathbf{r}-\mathbf{r}_{i,\alpha})$, where $u(\mathbf{r}-\mathbf{r}_{i,\alpha})$ is the Wannier function and $\mathbf{r}_{i,\alpha}$ ($\alpha=1,2,3$) are lattice vectors in the three sub-lattices of the $i$-th cell. Then the tight-binding Hamiltonian for our BEC system reads
\begin{equation}
\hat H_{TB}= \sum_{\langle i,j\rangle, \alpha,\beta} (t_{\alpha,\beta} \Psi_{i \alpha}^{*} \Psi_{j \beta}+\mathrm{H.c})+\sum_{i,\alpha}g|\Psi_{i \alpha}|^4\,,\label{eq_htb}
\end{equation}
where $\langle i,j\rangle$ runs over all nearest-neighboring sites, $\alpha/\beta$ includes the sublattices in one unit cell, $t_{\alpha,\beta}\simeq \int d\mathbf{r}[\frac{\hbar}{2m}(\nabla\Psi_{i \alpha}\cdot\nabla\Psi_{j \beta})+\Psi_{i \alpha}V_{latt}(\mathbf{r})\Psi_{j \beta}]$ is the hopping constant and $g$ is the nonlinear strength, for simplicity we take $t_{\alpha,\beta}=t$ and do not consider the next-nearest hopping in the main text. In experiments, one may upload $\mathrm{^{87}Rb}$ atoms which weigh $m=1.443\times 10^{-25}~\mathrm{kg}$ into the optical lattice. The lattice can be created by a laser of wavelength $\lambda=1064~\mathrm{nm}$. Thus the recoil energy is estimated by $E_R=\hbar^2k_L^2/2m \sim 2\pi\times 2 ~\mathrm{kHz}$. And the tunneling rate $t \sim 1~\mathrm{kHz}$ for a lattice depth $V=2 E_R$ in the absence of nonlinearity \cite{Miyake2013}.

In momentum space, $\hat H_{k}=\sum_{\mathrm{k} } \Psi_{\mathrm{k}}^{\dagger} \mathcal{H}_{\mathrm{k}} \Psi_{\mathrm{k} }$ with $\Psi_{\mathbf{k}}=\left(\Psi_{1 \mathbf{k} }, \Psi_{2 \mathbf{k}}, \Psi_{3 \mathbf{k}}\right)^{T}$, and
\begin{equation}
\mathcal{H}_{\mathrm{k}}=\hat H_0(\mathbf{k})+g\hat N=\mathbf{d}(\mathbf{k}) \cdot \mathbf{S}+g\hat N\,,\label{eq_TB}
\end{equation}
where $\mathbf{d}(\mathbf{k})=(d_x,d_y)$ denotes the Bloch vector with $d_x=2t\cos(k_x a/2)$ and $d_y=2t\cos(k_y a/2)$, and $\mathbf{S}=(S_x,S_y)$ are spin-one matrices with $S_x=\lambda_1$, $S_y=\lambda_6$ from the Gell-Mann matrices (for explicit form, see Appendix \ref{Appa}), together with $S_z=\lambda_5$ which consist the $\mathrm{SU(2)}$ Lie algebra $[S_{m}, S_{n}]=\mathrm{i} \epsilon_{m n \ell} S_{\ell}$. $\hat N=\operatorname{diag}(|\Psi_{1\mathbf{k}}|^{2},|\Psi_{2\mathbf{k}}|^{2},|\Psi_{3\mathbf{k}}|^{2})$ is the nonlinear term. The Bloch Hamiltonian $\mathcal{H}_k$ in absence of  nonlinear term $g \hat N$ has energy dispersions,
\begin{equation}
E_{\mathbf{k}}=0,\pm 2t\sqrt{\cos^2(k_x a/2)+\cos^2(k_y a/2)\,,}
\end{equation}
as illustrated in Fig. \ref{fig1} (c), which are in good agreement with Bloch bands of the continuous system Eq. (\ref{eq_GP}), see Appendix \ref{Appb}.

 \begin{figure*}[htbp]
	\centering
	\includegraphics[width=0.9\textwidth]{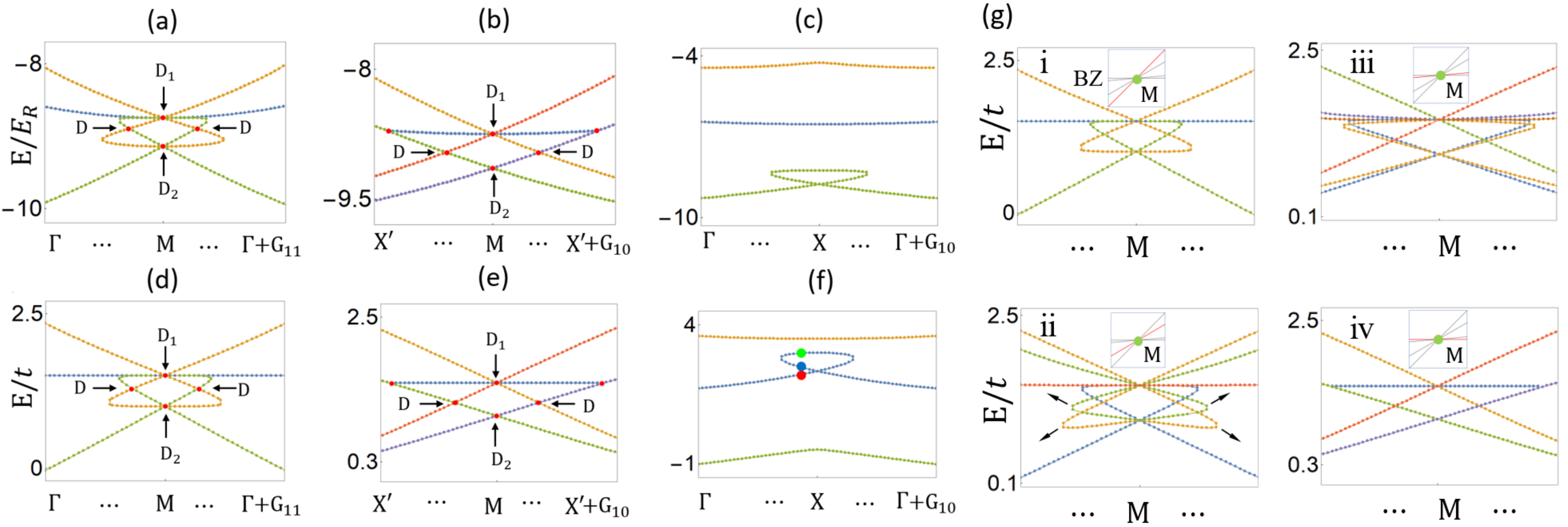}
	\caption{(a)-(c) Snapshots of energy bands for the continuous model Eq. (\ref{eq_GP}). (d)-(f) Snapshots of energy bands for the tight-binding model Eq. (\ref{eq_htb}). (g) Snapshots of energy bands for the tight-binding model along different sweeping paths across $M$. The energy units chosen are (a)-(c) the recoil energy $E_R=\hbar^2k_L^2/2m$ and (d)-(g) the hopping constant $t$ respectively. Parameters chosen are (a)-(c) $(V_{long},V_{short},V_{diag})=(4,4,4)E_R$, $U=(1.5,1.5,2.5)E_R$ respectively, and (d)-(g) $g=3 t$. The snapshots are taken within about $0.3 L_{\mathrm{BZ}}$ of width, where $L_{BZ}$ is the width of the first BZ along the $k_x$ ($k_y$) direction. The arrows in (g,ii) show how the wedged tube structure extends along the sweeping path. }
	\label{fig2}
\end{figure*}

Note that the low energy effective Hamiltonian $\mathcal{H}_0(\mathbf{q})$ has a $\hat C_4^z$ rotation symmetry $\hat C_4^z \mathcal{H}_0(\mathbf{q}) (\hat C_4^z)^{-1}=\mathcal{H}_0(D_{\hat C_4^z}\mathbf{q})$, where $\hat C_4^z \equiv e^{-i\frac{\pi}{2}S_z}$ and $D_{\hat C_4^z}\mathbf{q}=(-q_y,q_x)$ and a sublattice symmetry $\hat{O}\mathcal{H}_0(\mathbf{q})\hat{O}^{-1}=-\mathcal{H}_0(\mathbf{q})$, where $\hat{O}\equiv \operatorname{diag}(1,-1,1)$.

In fact the Lieb lattice has a $4mm$ plane point group symmetry. The quasi-momentum $\mathbf{k}$ is invariant with operations in little group $4mm$ at $M$ and  $2mm$ at $X$. And the invariant line $L_{\Sigma}$ and $L_{\Delta}$ are protected by mirror symmetry $m_d$ and $m_x$ respectively, thus indegenerate, while the Dirac point at M is protected by the 4mm symmetry featuring a nontrivial two dimensional irreducible representation. However, in presence of a nonzero nonlinear term $g\hat N$, even weak interaction will violate this symmetry, lifting the degeneracy at M and resulting in two more additional crossings, as shown in Fig. \ref{fig2}.
%the fragile symmetry will be broken by a nonlinear term $g\hat N$ even for the slightest interaction, lifting the degeneracy at M and resulting in two more additional crossings as shown in Fig. 2.
%While $4mm$ has a two dimensional irreducible representation that protects the Dirac point at $M$.
%A arbitrarily small $g\hat N$ breaks the $4mm$ symmetry,
%A nonlinear term $g\hat N$ even for arbitrary small interaction will break the $4mm$ symmetry, lifting the degeneracy at $M$ and results in two more additional crossings, as shown in Fig. \ref{fig2}.

We numerically calculate the Bloch bands  of the GP equation Eq. (\ref{eq_GP}) with the Bloch wave solutions  which are of the form $\psi_{\mathbf{k}}(\mathbf{r})=\sum_{m, n} c_{m n} e^{i\left(\mathbf{k}+\mathbf{G}_{m n}\right) \cdot \mathbf{r}}$, here $\mathbf{G}_{m n}=m \mathbf{b}_{1}+n \mathbf{b}_{2}$ is the reciprocal lattice vector. The results are shown in Fig. \ref{fig2}, which are in good agreement with that of the tight binding model $\mathcal{H}_{\mathrm{k}}$ in the vicinity of $M$. As predicted by previous works \cite{Chen2011}, the Bloch bands of a nonlinear two dimensional system possess tube structures and the Dirac point is extended into a closed self-crossing loop. The bands of our system have similar structures around the $M$ point. As illustrated in Fig. \ref{fig2} (g), the energy bands consist of three "tubes", which intersect at point $M$. They have wedged cross-section with size increasing  monotonically from $L_\Sigma$ to $\overline{L}_\Delta$ ($\overline{L}_\Delta$ denotes the invariant line cross $M$ and parallel to $k_x$-direction) as shown in Fig. \ref{fig2} (g), (\rmnum{2})-(\rmnum{3}), and overlap with each other to produce the structure of  Fig. \ref{fig2} (g), (\rmnum{1}/\rmnum{4}). For $g=0$ the Dirac point at $M$ accidentally degenerates with the flat band, which produces a triple point. As shown before, the Dirac point at $M$ will be lifted to two crossing points $D_1$ and $D_2$ for any $g \neq 0$ as the nonlinearity breaks the symmetry. However, the loop structures at the zone of Brillouin zone, as illustrated in Fig. \ref{fig2} (c) [or Fig. \ref{fig2} (f)], only emerges as the nonlinear strength $U$ ($g$) exceeds a critical value $U_c$ ($g_c$). Specifically, the nonlinearity does not modify the flat band on the invariant line $L_\Sigma$, as shown in Fig. \ref{fig2} (a) and (c). Furthermore, the flat bands of the tight binding model for $g=0$ and $g\neq 0$ on $L_\Sigma$ share the same eigen-states, only with a uniform shift of the eigen-value from $E=0$ to $E=g/2$ for any finite $g$.

%%%%%%%%%%%%%%%%%%%%%%%%%%%%%%
\section{Landau-Zener transition}\label{sec3}
Now we consider atoms performing Bloch oscillations with a constant applied force $\mathbf{F}$. As these atoms are accelerated in the vicinity of the triple Dirac point, the tunneling probability to other bands is finite.  This is a problem first considered by Landau and Zener, usually referred to as Landau-Zener (LZ) transition \cite{LandauZener1932,XTan2014,Shevchenko2010,Oliver2005,Car1986}.

 The effective Hamiltonian can be obtained from Eq. (\ref{eq_TB}) by performing Peierls substitution $\mathbf{k}=\mathbf{F}t$,
\begin{equation}
\hat{H}_{F}(t)=e^{-i t \boldsymbol{F} \cdot \hat{r}}(\hat{H}-i \partial_{t}) e^{i t \boldsymbol{F} \cdot \hat{r}}=\hat{H}(\boldsymbol{F} t)\,,
\end{equation}
where $\hat{H}=\mathcal{H}_{\mathrm{k}}-\boldsymbol{F} \cdot \hat{\boldsymbol{r}}$. The Peierls substitution can also be understood in a semiclassical way that the quasimomentums are subjected to a constant force $\dot{\mathbf{k}}=\mathbf{F}$ \cite{Lim2015}. An initial state $|\psi(t_0)\rangle=|u_n(\mathbf{k}_0)\rangle$ in the n-th band evolves as
\begin{equation}
i \partial_{t}|\psi[\mathbf{k}(t)]\rangle= \hat H_F[\mathbf{k}(t)]|\psi[\mathbf{k}(t)]\rangle\,,
\end{equation}
where $\mathbf{k}(t)=\mathbf{k}_{0}+\mathbf{F} t$. The force $\mathbf{F}=\dot{\mathbf{k}}$  linearly increases momentum $\mathbf{k}$ of atoms and leads to a non-adiabatic transition to other bands near the crossing (or anti-crossing) point. From the finial state $|\psi(t_f)\rangle$ the transition probability is defined as
\begin{equation}
P_{nm}=\left|\left\langle\psi\left(t_{f}\right) | u_{m}\left(\mathbf{k}_{f}\right)\right\rangle\right|^{2}\,,
\end{equation}
where $m,n=1,2,3$ and $| u_{m}(\mathbf{k}(t))$ is m-th eigen-state of $\hat H_F[\mathbf{k}_f]$, with $\mathbf{k}_f=\mathbf{k}_0+\mathbf{F} t_f$ at the final time $t_f$.
%%%%%%%%%%%%%%%%%%%%%%%%

\subsection{Adiabatic limit}

\begin{figure}
	\centering
	\includegraphics[width=\textwidth]{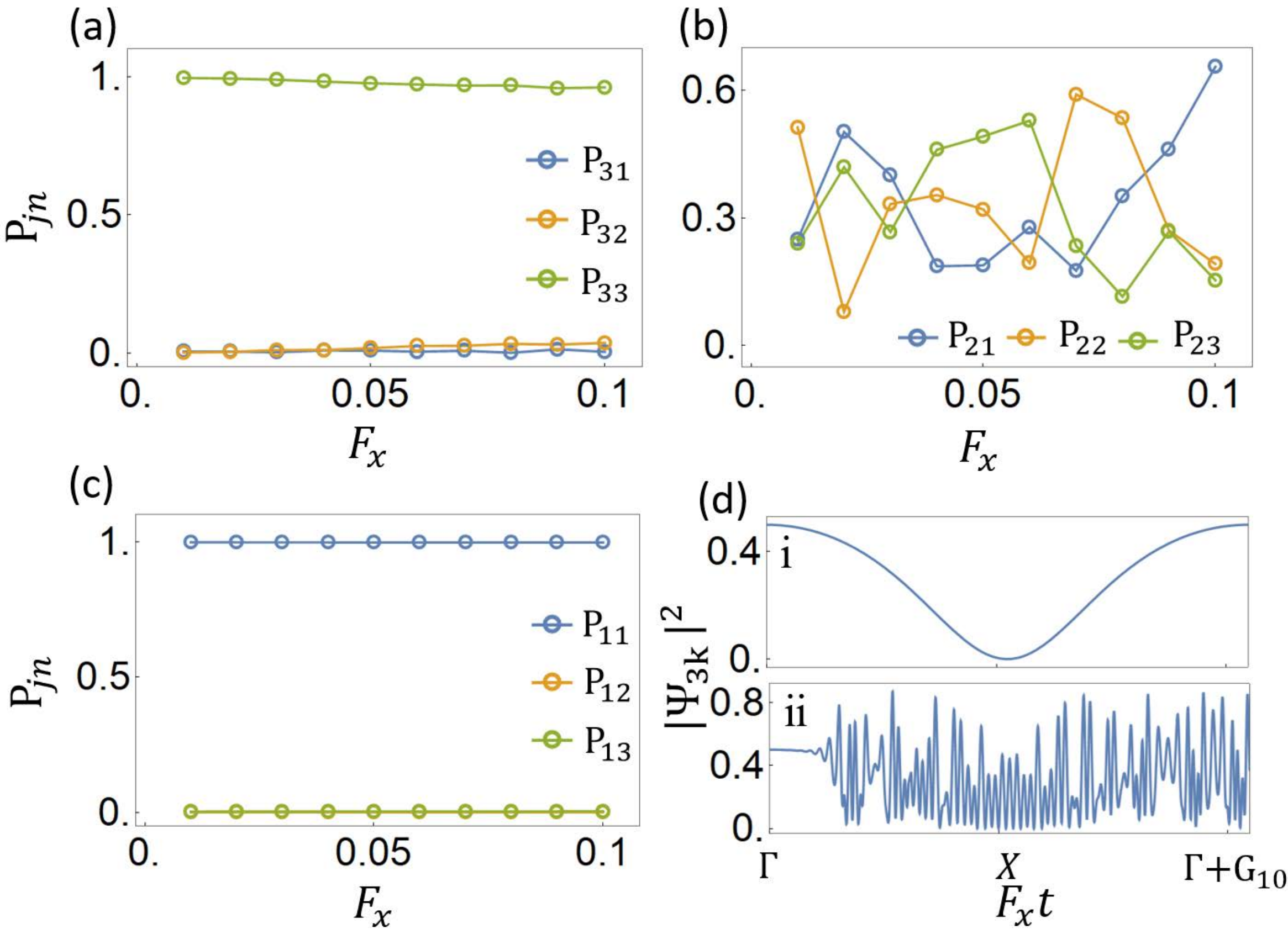}
	\caption{(a), (b) and (c) Transition probability for Landau-Zener tunneling in $k_x$ direction with $t=1$ and $g=3$. The atoms are initially prepared in (a) the up band, (b) the middle band and (c) the lowest band, respectively. (d) Time evolution of the particle number density of $3rd$ sublattice  $|\Psi_{3\mathbf{k}}|^2$ during the Landau-Zener process with $F=0.03$ for (\romannumeral1) g=0 and (\romannumeral2) g=3.  Here we have mapped the evolution time $t$ to the corresponding evolved momentum via the relation $k_x(t)=F_x t$. \label{fig3}}
\end{figure}

\begin{figure}[htbp]
	\centering
	\includegraphics[width=\textwidth]{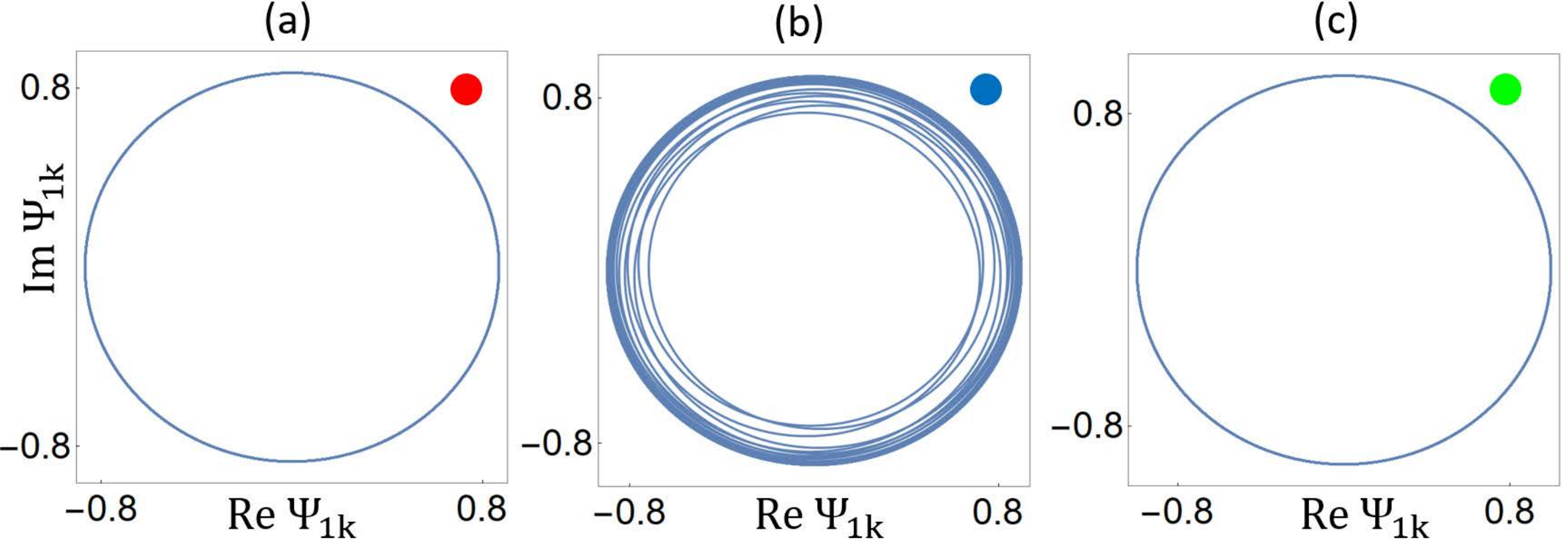}
	\caption{Time evolution of $\Psi_{1\mathbf{k}}$ for $i\partial_t \Psi_k=\mathcal{H}_k \Psi_k$ at $\mathbf{k}=(0.48\pi,0)$. The initial states are chosen as eigen-states of $\mathcal{H}_k$ located at the second band. The colored dots on the corner imply the location of eigenstates as depicted in Fig.\ref{fig2} (f).}
	\label{fig4}
\end{figure}

\begin{figure*}
	\centering
	\includegraphics[width=0.9 \textwidth]{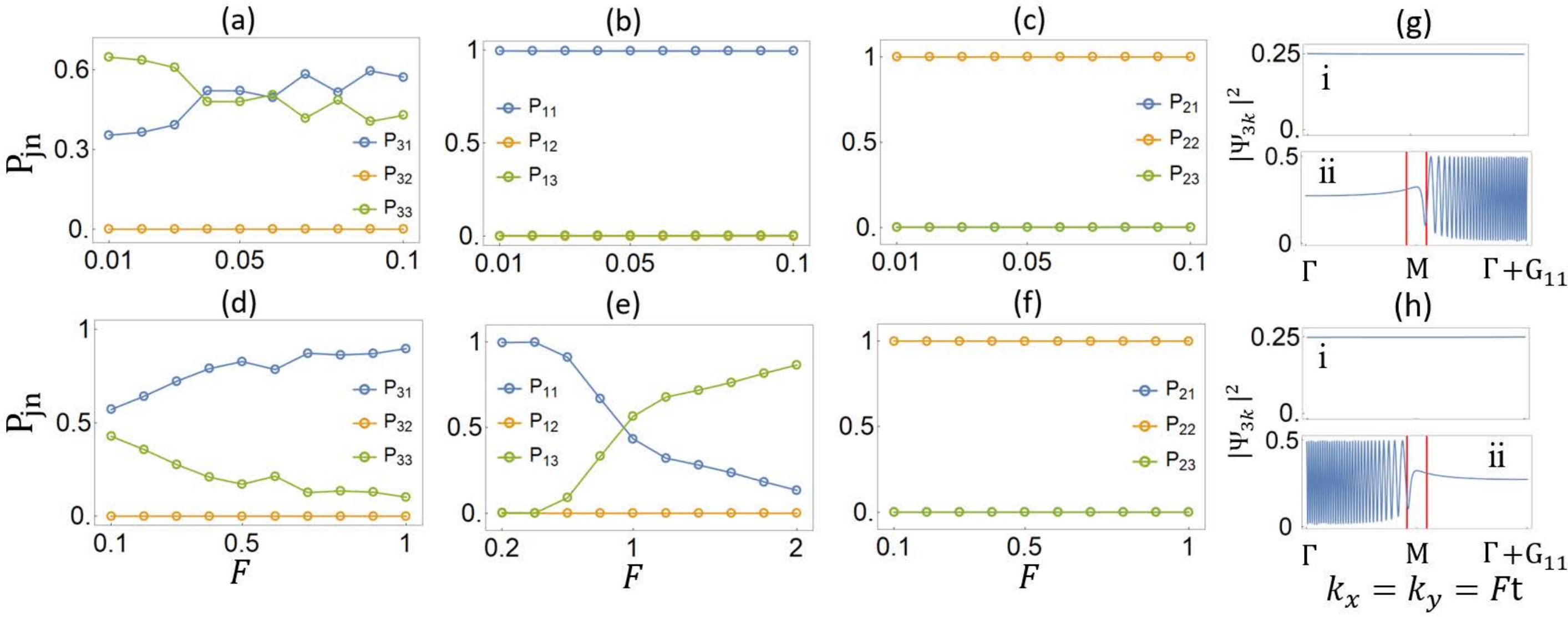}
	\caption{(a)-(f) Transition probability that depends on $\mathbf{F}$ for Landau-Zener tunneling along diagonal path $L_\Sigma$ [defined in Fig. \ref{fig1} (b)] with $t=1$ and $g=3$ for (a), (b) and (c) smaller force and (d), (e) and (f) larger force; (g) Time evolution of the particle number density of the $3rd$ sublattice $|\Psi_{3\mathbf{k}}|^2$ during the LZ process with $F=0.03$ for (\romannumeral1) g=0 and (\romannumeral2) g=3; (h) Time evolution of the particle number density of the 3rd sublattice $|\Psi_{3\mathbf{k}}|^2$ during the opposite sweeping for (\romannumeral1) g=0 and (\romannumeral2) g=3. The red lines in (g) and (h) denote the regime of tube structures.\label{fig5}}
\end{figure*}

For simplicity, we start with considering atoms performing one Bloch oscillation in $k_x$ directions. We numerically calculate the interband transition probabilities as a function of force $F_x$ (see Fig. \ref{fig3}). The applied force is given by $\mathbf{F}=F_x\hat e_x$ and atoms initially prepared in $\mathbf{k}=0$.  The evolution time is set to be $T=2\pi/(aF_x)$. Thus the momentum of atoms increases form $\mathbf{k}_0=0$ to $\mathbf{k}_f=2\pi/a$ under the influence of the force, along path of which the system has the spectrum shown in Fig. \ref{fig2} (f).

For g = 0, if the atoms take shorter time to travel across the anti-crossing than the  Zener tunneling time $T_{lz}=\sqrt{\overline\delta}\mathrm{max}(1,\sqrt{\overline\delta})/\Delta$, the system will evolve adiabatically, where $\overline\delta=\Delta^2/(4F_x)$ is the adiabaticity parameter and $\Delta$ is the minimal energy splitting \cite{Shevchenko2010}. For the parameters adopted in this paper, this corresponds to $F_x<1$. In presence of nonlinearity, the adiabaticity only holds for the bands whose structure is not modified by the nonlinearity. As seen in Fig. \ref{fig3} (a) and (c), the interband transition probabilities are vanishing while the force is weak enough. However, as shown in Fig.\ref{fig3} (b), the adiabaticity breaks down when atoms are initially prepared in the second band, since the nonlinearity changes the topological structure of this band.

Furthermore, the tunneling probability shows an irregular oscillation for atoms initially in the band with a loop structure, as observed in similar nonlinear systems studied in previous works \cite{Graefe2006,Wang2006}. This irregularity is thought to be associated with the chaotic behavior of the nonlinear system \cite{Wang2006}. As shown in Fig. \ref{fig3} (d), the evolution of state exhibits chaotic features, which could be directly observed in the laboratory by detecting the particle number density of the sublattices by means of hybrid time-of-flight images. For diagonal sweeping, the oscillation of state emerges shortly after the atoms travel across the band loops. Thus we could determine the regime of the fold in the energy bands.  In fact, the time evolution of $\mathcal{H}_k$ at the loops is unstable, which leads to the chaos in the LZ process. As illustrated in Fig. \ref{fig4}, the norm mode for the evolution of a eigen-state is cyclic. However, the periodicity breakdowns for the states on the fold, which implies the instability of the dynamics. The dynamical instability is attributed to the superfluidity of the BEC, which can be understood by the response of the superflow to a perturbation governed by the Bogoliubov equation \cite{WuL2001}, or the semiclassic fixed point theory of the Hamiltonian-Jacobi matrix \cite{Wang2006}. We adopt the latter method and give a brief introduction in Appendix \ref{appc}. As suggested by a previous work \cite{ZChen2010}, the Landau instability of superfluidity in a two-dimensional lattice is direction independent while the dynamical instability is not. Furthermore, Landau instability occurs beyond a critical wave number $\mathbf{k}_c$, and the superfluidity goes into a more stable phase as the  nonlinear interaction increases.

Next we consider tunneling process along the diagonal path $L_{\Sigma}$, \emph{i.e.}, the applied force is given by $\mathbf{F}=F_x\hat e_x+F_y\hat e_y$,  where we set $F_x=F_y=F$. The spectrum on this sweeping path has the structure shown in Fig. \ref{fig2} (d). The atoms are still initially prepared in $\mathbf{k}=0$ and the evolution time still set to be $T=2\pi/(aF_x)$. Fig. \ref{fig5} presents the transition probability as a function of the force $\mathbf{F}$. Different from the linear transition process, the relation of transition probability $P_{jn}=P_{nj}$ (see Appendix \ref{Appd}) does not hold for $g\neq 0$, since the nonlinear term $g\hat N$ breaks the particle-hole symmetry (or the symmetric structure of the spectrum). And particularly, as shown in Fig. \ref{fig5} (c) and (f), the results for atoms initially in the middle band remain the same as that of $g=0$, as the nonlinearity does not change the structure of the flat band.  Moreover, atoms prepared in the up or down band do not tunnel into the middle band when they travel across M, just like their behaviors in the absence of nonlinearity (see Appendix. \ref{Appd}).

\subsection{Sudden limit}
\begin{figure}[htbp]
	\centering
	\includegraphics[width=0.6\textwidth]{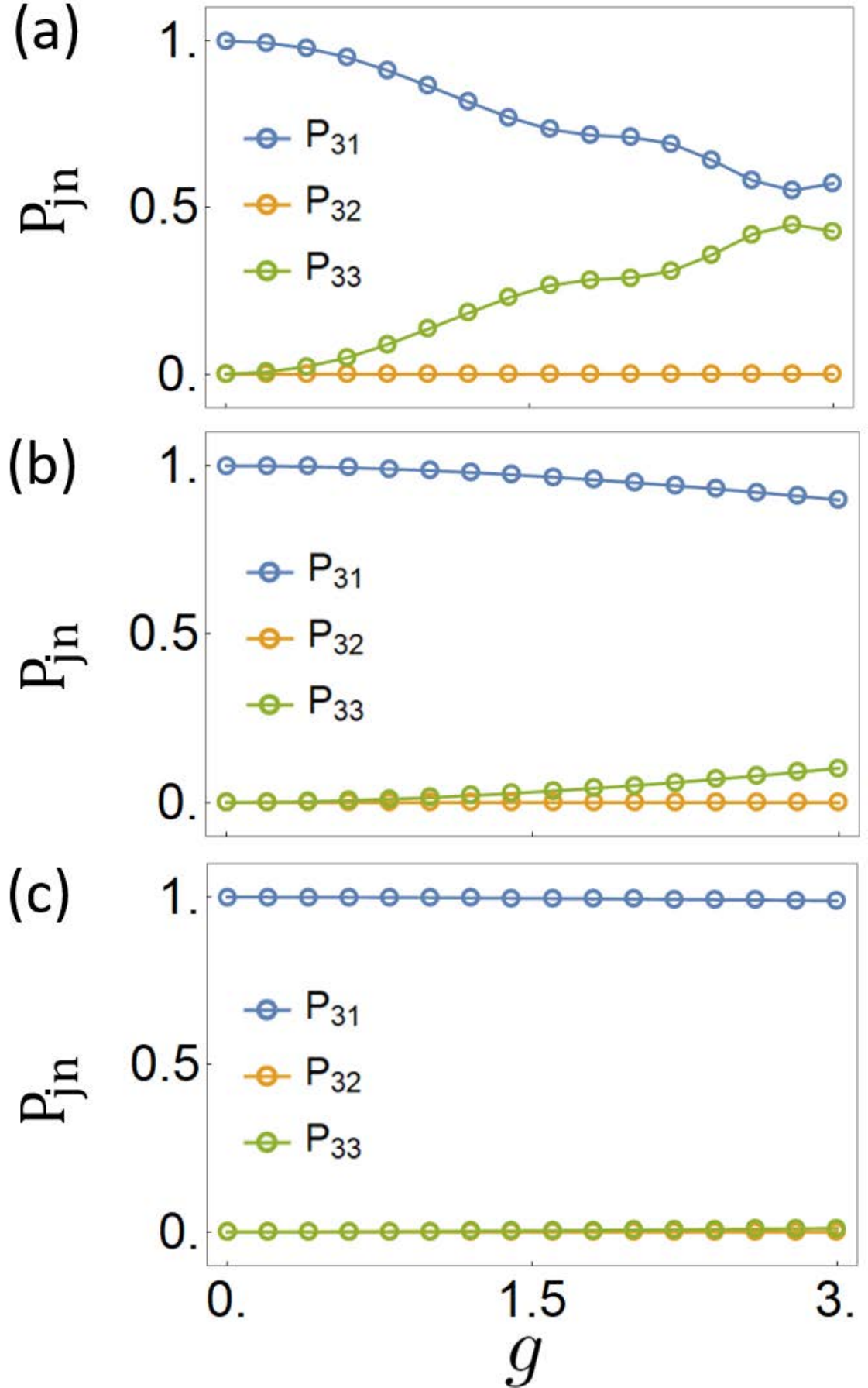}
	\caption{Transition probability that depends on the nonlinearity strength $g$ for Landau-Zener tunneling along diagonal path $L_\Sigma$ for (a) $F=0.1$ (b) $F=1.0$ (c) $F=10$.}
	\label{fig6}
\end{figure}
We consider the LZ transition acorss a single M point (gapless for $g=0$) in sudden limit, \emph{i.e.}, $F_x=F_y=F\gg 1$. As the force $F$ becomes large, the time that atoms take to travel across M point is much shorter than the Zener tunneling time. Thus the atoms stay in the initial state in most of the evolution time. The transition probability that depends on the nonlinear strength $g$ is displayed in Fig. \ref{fig6}. We find that difference between the transition probability for $g=0$ and $g\neq 0$ narrows as the sweeping velocity increases. And such difference is vanishing in the sudden limit, as shown in  Fig. \ref{fig6} (c), since the transition probability does not relate much to the structure of the spectrum. We could determine the nonlinearity $g$ of the condensate by fitting the experimental data with the interpolation function of $P_{jn}$ numerically.

\section{Conclusions}\label{sec4}

In conclusion, we have numerically studied the energy band structure of a BEC in an optical Lieb lattice and have provided a useful TB model. It is shown that under nonlinearity, the Dirac points at M point are broken and extended into a closed curve, while the flat band still exists on the high-symmetry line. We have shown that the Bloch-Zener oscillations could be utilized as a useful tool to explore the nonlinear bands: (\romannumeral1) the size of metastable fold bands can be determined by detecting the time evolution of the population of sublattices during Bloch oscillations, (\romannumeral2) the nonlinearity strength could be measured through the interband transition probability of specific tunneling process. Since the interatomic interaction can be easily manipulated, the observation of the phenomena studied here is expectable in the table-top ultracold atomic experiments.

%%%%%%%%%%%%%%%%%%%%%%
\acknowledgments

We thank Profs. Li-Jun Lang, Dan-Wei Zhang, and Shi-Liang Zhu for useful discussions. We also thank Zhu Chen for the help on numerical calculation. This work was supported by the Key-Area Research and Development Program of GuangDong Province (Grant No. 2019B030330001), the National Key Research and Development Program of China (Grant No. 2016YFA0301800), and the Key Project of Science and Technology of Guangzhou (Grant No. 201804020055), National Natural Science Foundation of China (Grants numbers 11704132), China Postdoctral Science Foundation (Grant number 2018M633063), and the Startup Foundation of SCNU.

%%%%%%%%%%%%%%%%%%%%%%
\begin{appendix}
\section{Mapping to universal Hamiltonian}\label{Appa}
The Gell-Mann matrices used in this paper is given by
\begin{equation}
\begin{split}
\lambda_{1}=\left[\begin{array}{lll}{0} & {1} & {0} \\ {1} & {0} & {0} \\ {0} & {0} & {0}\end{array}\right],\quad\lambda_{2}=\left[\begin{array}{ccc}{0} & {-i} & {0} \\ {i} & {0} & {0} \\ {0} & {0} & {0}\end{array}\right],\\
\lambda_{6}=\left[\begin{array}{ccc}{0} & {0} & {0} \\ {0} & {0} & {1} \\ {0} & {1} & {0}\end{array}\right],\quad\lambda_{7}=\left[\begin{array}{ccc}{0} & {0} & {0} \\ {0} & {0} & {-i} \\ {0} & {i} & {0}\end{array}\right].
\end{split}
\end{equation}
At $\boldsymbol{k}_{M}=(\pi, \pi)$, an expansion of $\mathcal{H}_k$ for small momentum shift $\boldsymbol{q}$ gives rise to an effective Hamiltonian which has the universal form
\begin{equation}
\mathcal{H}_\mathbf{q}=\mathbf{q} \cdot \mathbf{S}+g\hat N\,,
\end{equation}
where $\mathbf{q}=(q_x,q_y)$ denotes deviation of the Bloch wave number $\mathbf{q}$  away from $\boldsymbol{k}_{M}$. And by a simple replacement of $q_j=i\partial_j$ with $j=x,y$, we obtain the motion equation in real space,
\begin{equation}
i \partial_{t} \Psi=[\widetilde{\mathcal{H}}(i \partial_{j})+g \hat{N}] \Psi\,,
\end{equation}
where $\Psi=(|\Psi_{1}|^2,|\Psi_{2}|^2,|\Psi_{3}|^2)^T$ and $\widetilde{\mathcal{H}}(i \partial_{j})$ is $\mathcal{H}_\mathbf{q}$ with the replacement $q_j=i\partial_j$ ($j=x,y$).
%%%%%%%%%%%%%%%%%%%%
\section{NUMERICAL BAND STRUCTURE CALCULATION FOR $g=0$}\label{Appb}
\begin{figure}[htbp]
	\centering
	\includegraphics[width=\textwidth]{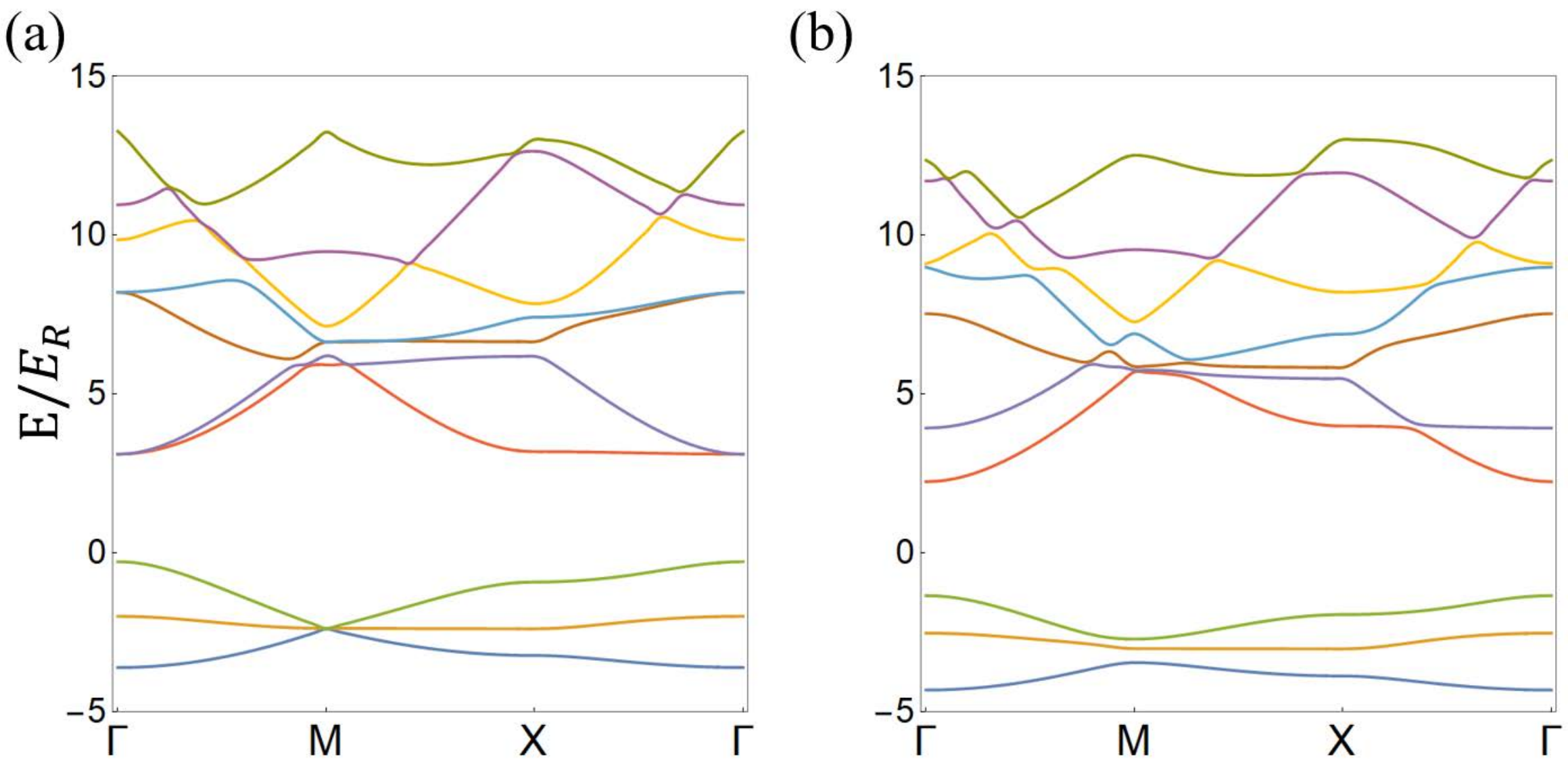}
	\caption{The lowest $10$ energy bands for $g=0$ with (a) $(V_{long}^{(x)},V_{short}^{(x)},V_{diag}^{(x)})
=(V_{long}^{(z)},V_{short}^{(z)},V_{diag}^{(z)})=(8,8,5.1)E_R$, and (b) $(V_{long}^{(x)},V_{short}^{(x)},V_{diag}^{(x)})
=(V_{long}^{(y)},1.5V_{short}^{(y)},0.8V_{diag}^{(y)})=(8,12,4)E_R$.}
	\label{figa1}
\end{figure}
We numerically study the band  structure \emph{ab initio} with the plane wave truncate approximation. We take Fourier transformation and write down the Hamiltonian in momentum space. The eigen-states have the form $\psi_{\mathbf{k}}(\mathbf{r})=\sum_{m, n} c_{m n} e^{i\left(\mathbf{k}+\mathbf{G}_{m n}\right) \cdot \mathbf{r}}$. And in reciprocal space the potential Eq. (\ref{eq_pot}) is writen as $V(\mathbf{r})=\sum_{m n} V_{m n} \exp \left(\mathbf{i} \mathbf{G}_{m n} \cdot \mathbf{r}\right)$, with components
\begin{equation}
\begin{split}
V_{m n}=&
-(0.25V_{long}^{(x)} (\delta _{m,-1}+2 \delta _{m,0}+\delta _{m,1}) \delta _{n,0}+\\&0.25 V_{long}^{(y)} \delta _{m,0} (\delta _{n,-1}+2 \delta _{n,0}+\delta _{n,1}))-\\&[0.25 V_{short}^{(x)}(\delta _{m,-2}+2 \delta _{m,0}+\delta _{m,2}) \delta _{n,0}+\\&0.25 V_{short}^{(y)} \delta _{m,0} (\delta _{n,-2}+2 \delta _{n,0}+\delta _{n,2})]-\\&[0.5 V_{diag}^{(x)} (-0.5 \delta _{m,1} \delta _{n,-1}+\delta _{m,0} \delta _{n,0}-0.5 \delta _{m,-1} \delta _{n,1})+\\&0.5 V_{diag}^{(y)} (-0.5 \delta _{m,-1} \delta _{n,-1}+\delta _{m,0} \delta _{n,0}-0.5 \delta _{m,1} \delta _{n,1})]\,.\end{split}
\end{equation}
Then by diagonalizing the Hamiltonian
\begin{equation}
\mathbf{H}^{\mathbf{k}}_{\left(m_{1}, n_{1};m_{2}, n_{2}\right)}=\delta_{m_{1}, m_{2}} \delta_{n_{1}, n_{2}} E_{\mathbf{k}+\mathbf{G}_{m_{1}, n_{1}}}^{0}+V_{m_{1},m_{2}, n_{1},n_{2}}\,,
\end{equation}
we obtain the eigen-energies for a specific $\mathbf{k}$, where $E_{\mathrm{q}}^{0}=\hbar^{2} q^{2} / 2 m$ denotes the kinetic energy.

The results are shown in Fig. \ref{figa1}. Here we choose recoil energy $E_R=\hbar^2k_L^2/2m$ as the energy unit. The lowest three bands are quite well approximated by the tight binding model. Bands of Lieb lattice can be gapped for  some lattice configurations $(V_{long}^{(x,z)},V_{short}^{(x,z)},V_{diag}^{(x,z)})$.

%%%%%%%%%%%%%%%%%%%%%%%%%%
\section{Mapping to classical equations}\label{appc}

\begin{figure}[htbp]
	\centering
	\includegraphics[width=\textwidth]{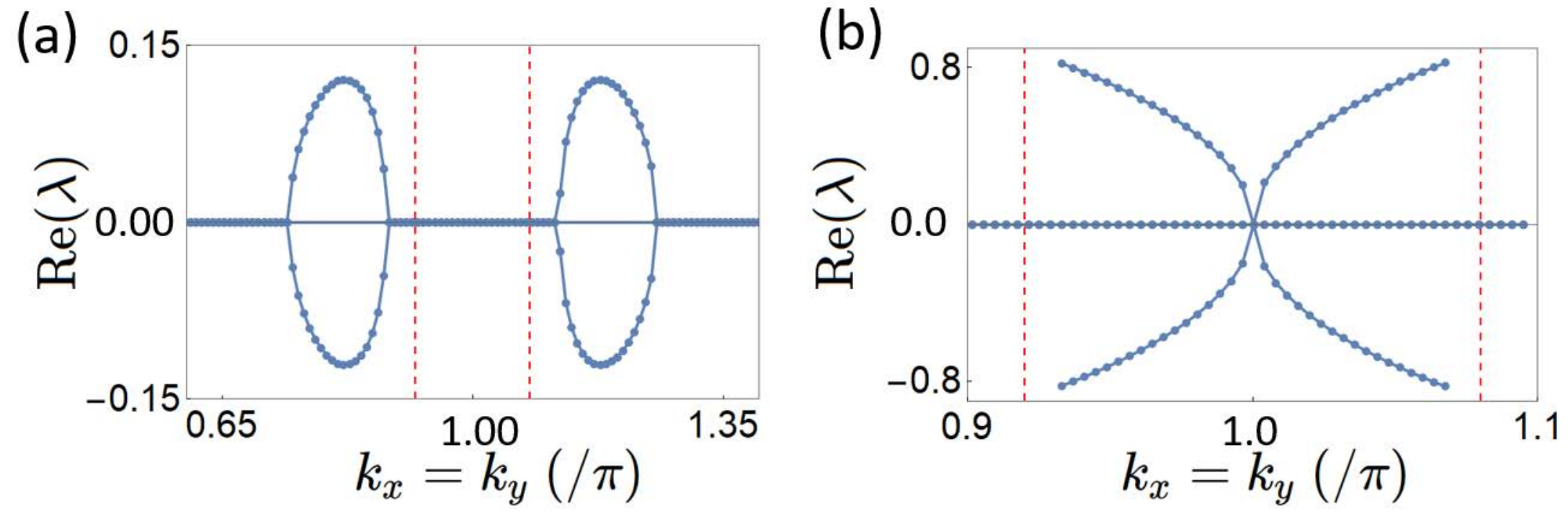}
	\caption{The real parts of eigenvalues of $H_J$ along $L_\Sigma$. The full values in the regime between the two dashed line in (a) are given in (b).}
	\label{figappc}
\end{figure}
The nonlinear three-level model Eq. \ref{eq_TB} can also be described as a classical Hamiltonian system, similar to Ref. \cite{Wang2006,Graefe2006}. With $\Psi_{1\mathbf{k}}=\sqrt{s_1}e^{i\theta_1}$,  $\Psi_{2\mathbf{k}}=\sqrt{1-s_1-s_2}e^{i\theta_2}$, $\Psi_{3\mathbf{k}}=\sqrt{s_2}e^{i\theta_3}$, the nonlinear TB Hamiltonian can be cast into a classical Hamiltonian system,
\begin{equation}
\begin{aligned}
\mathcal{H}=&\frac{g}{2}[s_{1}^{2}+s_{2}^{2}+(1-s_{1}-s_{2})^{2}]
+2 \sqrt{1-s_{1}-s_{2}}\times\\
&[\cos(k_x) \sqrt{s_1} \cos (\tilde{\theta}_{1})+
\cos(k_y) \sqrt{s_{2}} \cos(\tilde{\theta}_{2})]\,,
\end{aligned}
\end{equation}
where $\tilde{\theta}_1=\theta_1-\theta_2$,  $\tilde{\theta}_2=\theta_3-\theta_2$ are relative phase, and we treat $k_x$ and $k_y$ as parameters. $s1,\tilde{\theta}_1$ and $s2,\tilde{\theta}_2$ are two pairs of canonically conjugate variables which are governed by,
\begin{equation}
\dot{s}_{1}=-\cos(k_x) \sqrt{\left(1-s_{1}-s_{2}\right) s_{1}} \sin(\tilde{\theta}_{1})\,,
\end{equation}
\begin{equation}
\begin{aligned}
\dot{\tilde{\theta}}_{1}=&g(1-2 s_{1}-s_{2})-\frac{1-2 s_{1}-s_{2}}{2 \sqrt{\left(1-s_{1}-s_{2}\right) s_{1}}} \cos(k_x) \cos(\tilde{\theta_{1}}) \\
&+\frac{s_{2}}{2 \sqrt{(1-s_{1}-s_{2}) s_{2}}} \cos(k_y) \cos (\tilde{\theta}_{2})\,,
\end{aligned}
\end{equation}
\begin{equation}
\dot{s}_{2}=-\cos(k_y) \sqrt{\left(1-s_{1}-s_{2}\right) s_{2}} \sin(\tilde{\theta}_{2})\,,
\end{equation}
\begin{equation}
\begin{aligned}
\dot{\tilde{\theta}}_{2}=&g(1-2 s_{1}-s_{2})+\frac{s_{1}}{2 \sqrt{\left(1-s_{1}-s_{2}\right) s_{1}}} \cos(k_x) \cos(\tilde{\theta_{1}}) \\
&-\frac{1-s_1-s_{2}}{2 \sqrt{(1-s_{1}-s_{2}) s_{2}}} \cos(k_y) \cos (\tilde{\theta}_{2})\,.
\end{aligned}
\end{equation}
By setting $\dot{s}_{1}=\dot{s}_{2}=\dot{\theta}_{1}=\dot{\theta}_{2}=0$, the eigenenergy is obtained from $\epsilon=\mathcal{H}$. And the chaotic nature of the nonlinear system can be revealed by the Hamiltonian-Jacobi matrix, which is obtained by linearizing the above motion equations at the fixed points (corresponding to the eigenstates),
\begin{equation}
H_{J}=\left[\begin{array}{cccc}
-\frac{\partial^{2} H_{e}}{\partial s_{1} \partial \theta_{1}} & -\frac{\partial^{2} H_{e}}{\partial^{2} \theta_{1}} & -\frac{\partial^{2} H_{e}}{\partial s_{2} \partial \theta_{1}} & -\frac{\partial^{2} H_{e}}{\partial \theta_{2} \partial \theta_{1}} \\
\frac{\partial^{2} H_{e}}{\partial^{2} s_{1}} & \frac{\partial^{2} H_{e}}{\partial \theta_{1} \partial s_{1}} & \frac{\partial^{2} H_{e}}{\partial s_{2} \partial s_{1}} & \frac{\partial^{2} H_{e}}{\partial \theta_{2} \partial s_{1}} \\
-\frac{\partial^{2} H_{e}}{\partial s_{1} \partial \theta_{2}} & -\frac{\partial^{2} H_{e}}{\partial \theta_{1} \partial \theta_{2}} & -\frac{\partial^{2} H_{e}}{\partial s_{2} \partial \theta_{2}} & -\frac{\partial^{2} H_{e}}{\partial^{2} \theta_{2}} \\
\frac{\partial^{2} H_{e}}{\partial s_{1} \partial s_{2}} & \frac{\partial^{2} H_{e}}{\partial \theta_{1} \partial s_{2}} & \frac{\partial^{2} H_{e}}{\partial^{2} s_{2}} & \frac{\partial^{2} H_{e}}{\partial \theta_{2} \partial s_{2}}
\end{array}\right]\,.
\end{equation}
We numerically solve the eigenvalues $\lambda$ of $H_J$ along the invariant line $L_\Sigma$. The results are shown in Fig. \ref{figappc}. In general, the eigenvalues are a complex number and only solutions with pure imaginary values are stable. We can see that near $M$ point the real parts of the eigenvalue become nonzero, which implies the emergency of the unstable fold structure.
%%%%%%%%%%%%%%%%%%%%%%%
%%%%%%%%%%%%%%%%%%%%%%%
\section{Landau-Zener transition for $g=0$}\label{Appd}
\begin{figure}[htbp]
	\centering
	\includegraphics[width=0.8\textwidth]{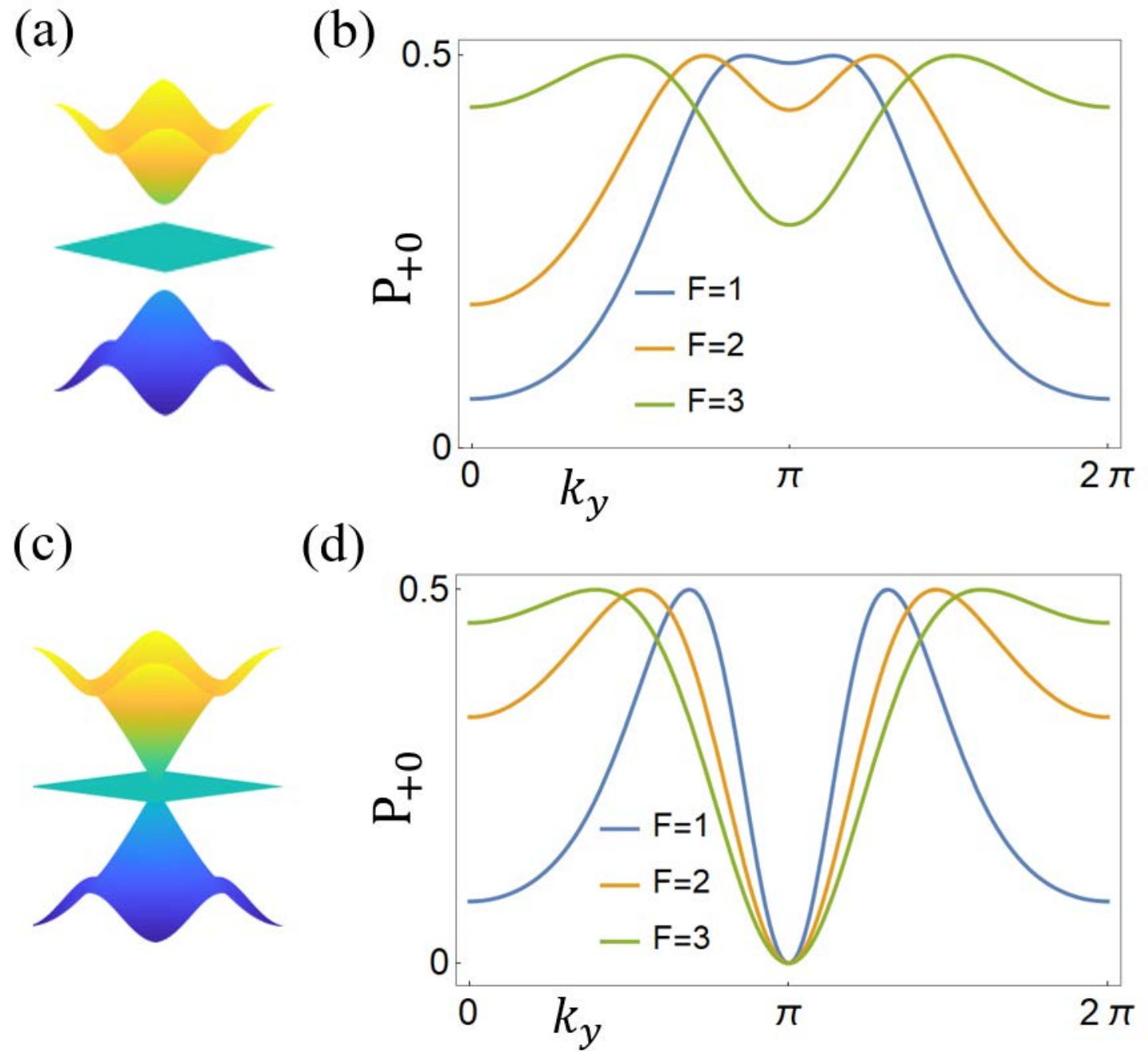}
	\caption{(a) and (c) Energy spectrum for $\delta=0.3$ and $\delta=0$. (b) and (d) Transition probability $P_{+0}$ with a constant force in $k_x$ direction for $\delta=0.3$ and $\delta=0$.}
	\label{figa2}
\end{figure}
We consider the tight binding Hamiltonian Eq. (\ref{eq_htb}) with $t_{i,1(3);i,2}=(1+\delta)t$ and $t_{i,1(3);i+1,2}=(1-\delta)t$ which possess a finite gap in the system . In momentum space,
\begin{equation}
\begin{split}
\mathcal{H}_{\mathbf{k}}=&2t\cos(k_x a/2)\lambda_1-2t\delta\sin(k_x a/2)\lambda_2+\\&2t\cos(k_y a/2)\lambda_6-2t\delta\sin(k_y a/2)\lambda_7\,.
\end{split}
\end{equation}
And the energy spectrum is given by
\begin{equation}
\epsilon(\mathbf{k})=0,\pm 2 t \sqrt{1+\delta^{2}+\left(1-\delta^{2}\right)\left(\cos k_{x} a+\cos k_{y} a\right) / 2}\,.
\end{equation}
Without loss of generality, we consider the motion along the $k_x$ direction. The constant force is given by $F_x=k_{x0}+Ft$, and the evolution time is chosen to be $t\in[0,2\pi/(a F)]$. Following the calculations outlined in Ref.\cite{Car1986,Wang2006}, we can obtain the transition probability for $\mathcal{H}_q$,
\begin{equation}
P_{++}=[1-\exp (-\frac{\pi \Delta^{2}}{4 F})]^{2}\,,
\end{equation}
\begin{equation}
P_{+0}=2 \exp (-\frac{\pi \Delta^{2}}{4 F})[1-\exp (-\frac{\pi \Delta^{2}}{4 F})]\,,
\end{equation}
\begin{equation}
P_{+-}=\exp (-\frac{\pi \Delta^{2}}{2 F})\,,
\end{equation}
\begin{equation}
P_{00}=[1-2 \exp(-\frac{\pi \Delta^{2}}{4 F})]^{2}\,,
\end{equation}
where $P_{jn}$ is the occupation probability on the $j$ band at $t=T_f$ for the state initially on the n band at $t=T_0$, and $\Delta=2 \sqrt{\frac{1}{2} \left(1-\delta^2\right) (\cos (k_y)-1)+\delta^2+1}$. And we have $P_{jn}=P_{nj}$ due to the symmetry of the levels.

We calculate the transition probabilities for $g=0$ with atoms performing a single Bloch oscillation in $k_x$ directions. The results are shown in Fig. \ref{figa2}. Since the parallel momentum $k_y$ is conserved, the term $2t\cos(k_y a/2)S_y$ can be seen as an effective mass term. Thus $P_{jn}$ is modified with different initial momentum $k_{y0}$. In particular, for system in gappless phase, \emph{i.e.}, $\delta=0$, and atoms prepared in initial momentum $\mathbf{k}=0$, the travel across the triple Dirac point leads to a unit transferred probability $P_{+-}=0$ ($P_{+0}=0$) as shown in Fig. \ref{figa2} (d).

\end{appendix}

%\bibliography{ref}
\end{document}